\documentstyle[11pt,paspconf,psfig]{article}
\def\refis{\ }

\def\gtorder{\mathrel{\raise.3ex\hbox{$>$}\mkern-14mu
             \lower0.6ex\hbox{$\sim$}}}
\def\ltorder{\mathrel{\raise.3ex\hbox{$<$}\mkern-14mu
             \lower0.6ex\hbox{$\sim$}}}

\begin{document}

%HERNQUIST}\footnote{$^\dagger$} {Alfred P. Sloan Foundation Fellow, 
%Presidential Faculty Fellow}}

\title{Triaxial Halos and Cusps
    }

\author{Steinn Sigurdsson$^1$, Chris Mihos$^2$, Lars Hernquist$^3$ \& Colin Norman$^4$}
\affil{$^1$,IoA, Cambridge, CB3 0HA, UK {\sl (EU Marie Curie Fellow)}; 
$^2$Dept. Ast., CWRU, Cleveland, OH 44106 {\sl (Hubble Fellow)}; 
$^3$BoAA, UCSC, Santa Cruz CA 95064;  
$^4$Dept. Phys. Astr., JHU, Baltimore, MD 21218   
    }

\begin{abstract}
We present $N$--body models for triaxial elliptical galaxies
or halos of galaxies, which are fully self--gravitating, have
near constant axis ratio as a function of radius and a
$r^{-1}$ central density cusp. Preliminary investigation suggests
the model are stable and orbit analysis shows no indication
of chaotic orbits.
The models provide a starting point for investigations
into the evolution of triaxial figures of equilibrium, response of
triaxial figures to central black holes, external perturbations
and interactions.
\end{abstract}
\par\noindent There is strong evidence that at least some galaxies
are triaxial (1,2,3).
Models known as ``perfect ellipsoids'' have been developed 
to describe elliptical galaxies (4),
but these have flat central cores, unlike those observed in
real ellipticals.
A problem of interest is whether it is possible to construct
self-gravitating spheroids that are stationary, cuspy, and triaxial,
and if so, how strong the cusp can be, and how strong the
triaxiality may be. It is also interesting to explore how
rapidly unstable figures evolve towards axisymmetry.
Clearly in the limit of weak cusps and in the limit
of weak triaxiality the figure evolution must either be very slow
or non--existent.
It has been conjectured that central singularities in the
potential or density (5,6,7,8) introduce chaotic orbits
that rapidly drive a secular evolution in figure shape.
Open questions include whether non--regular
orbits in non--integrable potentials are strongly chaotic or
only ``semi--stochastic'', and what
fraction of phase space is occupied by irregular orbits (5,9).

Most investigations of these issues have used either rigid potentials
or low resolution numerical models generated by cold collapse
from initially triaxial conditions. We describe here a new
method for generating self--gravitating triaxial spheroids
with high numerical resolution.
The models start off as spherical, with some initial density cusp,
$\rho(r) \propto r^{-\gamma}$. The models are realised as large $N$
particle distributions, using a multi--mass realisation (10).
The configurations are integrated using the SCF scheme (11).
The large number of particles and the large number of dynamical
times required makes this a problem particularly well suited
for parallel implementations of the SCF method (12,13). Our code uses
an adaptive Hermite integration scheme with variable multi--level
timestepping (Sigurdsson et al; Mihos et al; in preparation). 

\begin{figure}
\centerline{\psfig{figure=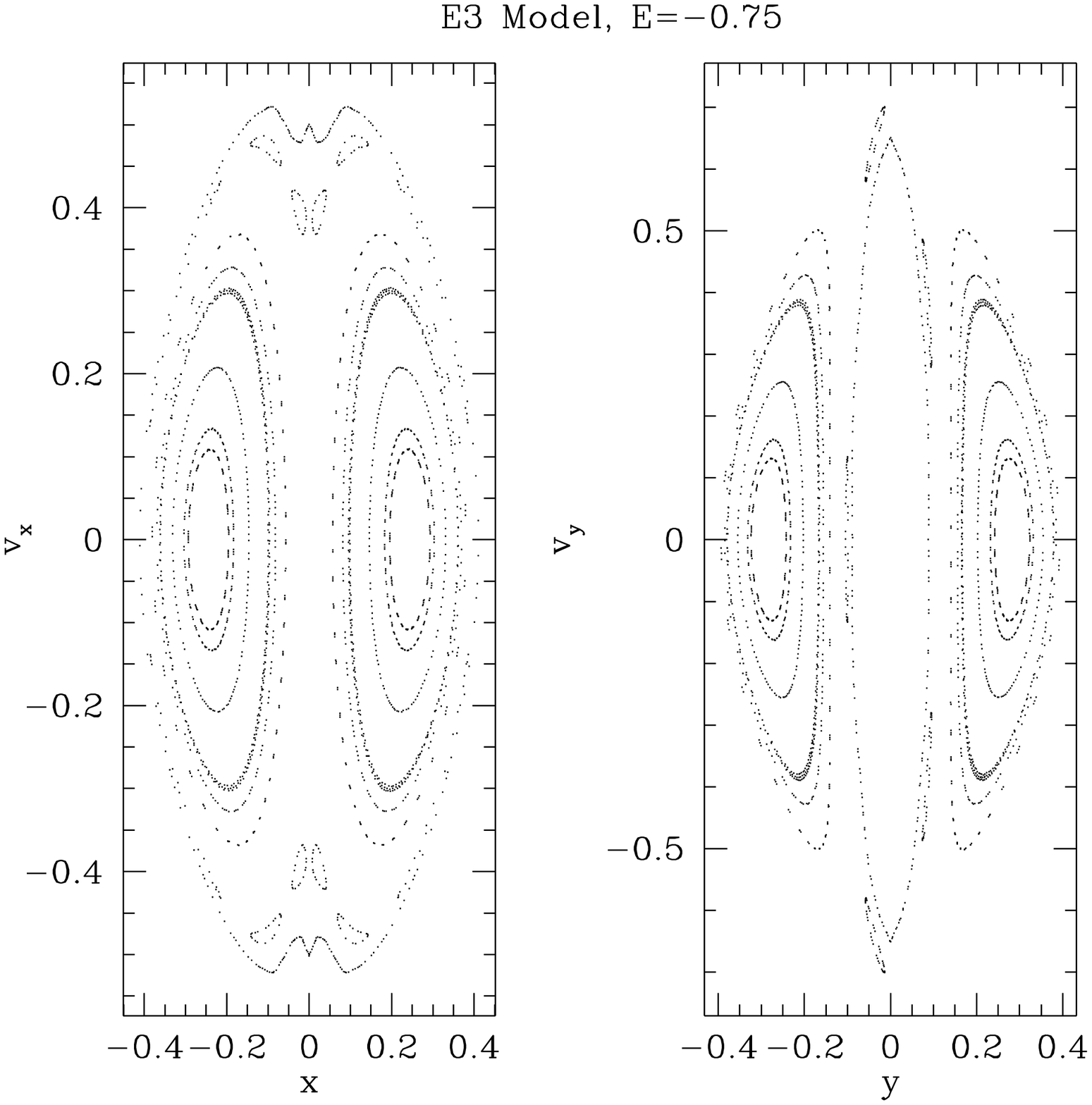,height=3.in,width=4.5in}}
\caption{
$E=-0.75$ surface of section. 
}
\end{figure}

To generate triaxial models, we squeeze the velocity ellipsoid
using a simple iterative scheme; 
$v_{y,z,i}(t) = v_{y,z,i}(t)(1 - \epsilon (dt,t,y,z))$.
The choice of parameters is such that over many dynamical timescales
the model settles to triaxial figure with ellipticities in
the range 0.1--0.5, and triaxiality parameter $T \sim 0.2 - 0.8$, 
with the axis ratios
close to constant with radius. The model discussed here has a density
cusp $\gamma = -1$, intermediate axis $b=0.85$ and minor axis $c=0.7$,
well approximated by a non--spherical Hernquist model.
To check for long term stability, 
the model is then evolved for tens of dynamical times and the figure
shape is compared with the initial shape. We see no evidence for secular
evolution in axis ratios, or the slope of the inner density cusp.
To investigate the intrinsic structure of the model we
have generated surfaces of section for the major axis
orbits. To do this, we freeze the potential
generated by the distribution after the model has relaxed, and
extract the SCF potential coefficients. These are then used to
generate a rigid potential for integrating test particle orbits
drawn either from the particles generating the density distribution
in the model, or by placing test particles at chosen points in phase space.
To check for the effect of numerical noise due to discreteness,
we smoothed the final potential using the reflection symmetry of
the distribution (a ``quiet finish'', analogous to the ``quiet start''
technique advocated for numerical studies of stability). While the
orbits in the unsmoothed potential are clearly noisier than in
the smoothed potential there is no apparent qualitative change in
orbit structure.

Figure 1 shows a surface of section for a set of particles
in the x--y plane (the x--axis being the major axis, the y--axis 
the intermediate axis, by designation) with energy $E=-0.75$
(the particle energies range from $0$ to $-1$).
Figure 2 shows a second surface of section for $E=-0.9$.
At the lower energy a large fraction of phase space is occupied by
boxlets, with the resonant orbits contributing to the global
figure shape as the box orbits break up. Some of the loops show
signs of high order resonances. Neither figure shows any indication of
chaotic orbits. 
\begin{figure}
%\centerline{\psfig{figure=fig2.ps,width=3.in}}
\centerline{\psfig{figure=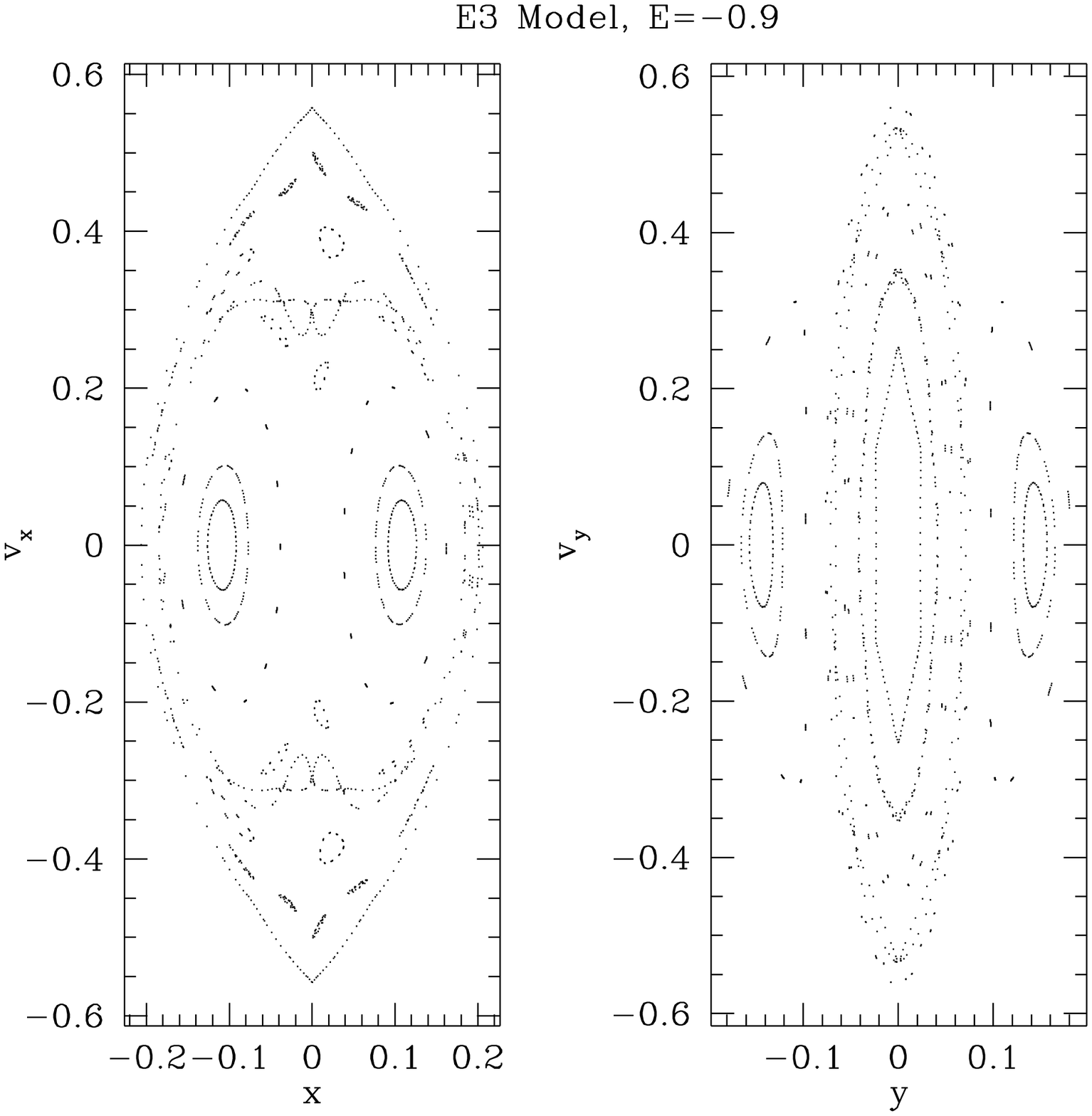,height=3.in,width=4.5in}}
\caption{
$E=-0.9$ surface of section.
}
\end{figure}

Clearly, moderately triaxial galaxies can exist with density
cusps at least as strong as $r^{-1}$.  It is possible that the
technique we use to generate the models ``traps'' a fraction of
the particles in resonant orbits which serve to support the
triaxial figure where other numerical techniques select for more
axisymmetric figures. The particular model shown here is one
of a family we have generated. Further analysis of the structure
of the models is continuing and we are using the models for
an investigation into the response of triaxial galaxies
to the presence of central black holes (in preparation).

\acknowledgments
This work was supported in part by an allocation of Cray T3E time from the
Pittsburgh Supercomputing Center, NASA under grants NAG5-3059 and NAG 5-3820,
NSF under grant ASC 93-18185, and the British Council.
\par\noindent
%\begin{references}
[1] de Zeeuw, T., Franx, M., 1991, ARAA, 29, 239 \\
\refis [2] Franx, M., Illingowrth, G., de Zeeuw, T., 1991, ApJ, 383, 112 \\
\refis [3] Statler, T.S., 1994, ApJ, 425, 458 \\
\refis [4] de Zeeuw, T., 1985, MNRAS, 216, 273 \\
\refis [5] Goodman, J., Schwarzschild, M., 1981, ApJ, 245, 1087 \\
\refis [6] Gerhard, O.E., Binney, J., 1985, MNRAS, 216, 467 \\
\refis [7] Norman, C.A., May, A., van Albada, T., 1985, ApJ, 296, 20 \\
\refis [8] Merritt, D., Fridmann, T., 1996, ApJ, 460, 136 \\
\refis [9] Miralda--Escude, J., Schwarzschild, M., 1989, ApJ, 339, 752 \\
\refis [10] Sigurdsson, S., Hernquist, L., Quinlan, G.D., 1995, ApJ, 446, 75 \\
\refis [11] Hernquist, L., Ostriker, J.P., 1992, ApJ, 386, 375 \\
\refis [12] Hernquist, L., Sigurdsson, S., Bryan, G.L., 1995, ApJ, 446, 717 \\
\refis [13] Sigurdsson, S., He, B., Melhem, R., Hernquist, L., 1997a, CiP, 11.4, 378 \\
%\end{references}

\end{document}